\documentclass[prx,aps,twocolumn,superscriptaddress,nofootinbib]{revtex4-2}

\usepackage{graphicx}% Include figure files
\usepackage{dcolumn}% Align table columns on decimal point
\usepackage{bm}% bold math
\usepackage{hyperref}% add hypertext capabilities
\usepackage{amsbsy,amssymb,amsmath,bm,mathtools}

\usepackage{cleveref}
\usepackage{natbib}
\usepackage{xcolor}

\newcommand{\R}{\ensuremath{\mathbb{R}} }
\DeclareMathOperator{\supp}{supp}

\begin{document}

%\preprint{APS/123-QED}

\title{Convolutional neural networks for large-scale dynamical modeling of itinerant magnets}
%\title{Acceleration of Spin Force Calculation using Artificial Intelligence}% Force line breaks with \\
%\thanks{A footnote to the article title}%

\author{Xinlun Cheng}
\affiliation{School of Data Science, University of Virginia, Charlottesville, VA 22903, USA}
\affiliation{Department of Astronomy, University of Virginia, Charlottesville, Virginia 22904, USA}

\author{Sheng Zhang} 
\affiliation{Department of Physics, University of Virginia, Charlottesville, Virginia 22904, USA}

\author{Phong C. H. Nguyen}
\affiliation{School of Data Science, University of Virginia, Charlottesville, VA 22903, USA}

\author{Shahab Azarfar}
\affiliation{School of Data Science, University of Virginia, Charlottesville, VA 22903, USA}

\author{Gia-Wei Chern} 
\affiliation{Department of Physics, University of Virginia, Charlottesville, Virginia 22904, USA}

\author{Stephen S. Baek}
\affiliation{School of Data Science, University of Virginia, Charlottesville, VA 22903, USA}
\affiliation{Department of Mechanical and Aerospace Engineering, University of Virginia, Charlottesville, VA 22903, USA}

\date{\today}% It is always \today, today,
             %  but any date may be explicitly specified

\begin{abstract}
Complex spin textures in itinerant electron magnets hold promises for next-generation memory and information technology. The long-ranged and often frustrated electron-mediated spin interactions in these materials give rise to intriguing localized spin structures such as skyrmions.  Yet, simulations of magnetization dynamics for such itinerant magnets are computationally difficult due to the need for repeated solutions to the electronic structure problems. We present a convolutional neural network (CNN) model to accurately and efficiently predict the electron-induced magnetic torques acting on local spins. Importantly, as the convolutional operations with a fixed kernel (receptive field) size naturally take advantage of the locality principle for many-electron systems, CNN offers a scalable machine learning approach to spin dynamics. We apply our approach to enable large-scale dynamical simulations of skyrmion phases in itinerant spin systems. By incorporating the CNN model into Landau-Lifshitz-Gilbert dynamics, our simulations successfully reproduce the relaxation process of the skyrmion phase and stabilize a skyrmion lattice in larger systems. The CNN model also allows us to compute the effective receptive fields, thus providing a systematic and unbiased method for determining the locality of the original electron models. 
\end{abstract}

%\keywords{Suggested keywords}%Use showkeys class option if keyword
                              %display desired
\maketitle

%\tableofcontents

\section{\label{sec:intro}Introduction}

Itinerant frustrated magnets with electron-mediated spin-spin interactions often exhibit complex non-collinear or non-coplanar spin textures. Of particular interest are particle-like objects such as magnetic vortices and skyrmions which are not only of fundamental interest in magnetism but also have important technological implications in the emerging field of \textit{spintronics}~\cite{bogdanov89,rossler06,muhlbaure09,yu10,yu11,seki12,nagaosa13}. These nanometer-sized localized spin textures are characterized by nontrivial topological invariants and are rather stable objects with long lifetimes. In itinerant electron magnets, skyrmions can be moved, created, and annihilated not only by magnetic fields but also by electrical currents thanks to electron-spin interactions. The presence of such complex textures could also give rise to intriguing electronic and transport properties, such as the topological Hall effects and topological Nernst effects~\cite{nagaosa13,loss92,ye99,tatara02}, due to a nontrivial Berry phase acquired by electrons when traversing over closed loops of non-coplanar spins~\cite{berry84}. 

%Also importantly, such topological electronic responses in metallic magnets can be controlled via the manipulation of magnetic textures.

Dynamical modeling of complex textures in itinerant spin systems, however, is a computationally challenging task. While magnetic moments in most metallic skyrmion materials can be well approximated as classical spin vectors, the local effective magnetic fields, analogous to forces in molecular dynamics, originate from exchange interactions with itinerant electrons and must be computed quantum mechanically. Dynamics simulations of such itinerant magnets thus require solving an electronic structure problem associated with the instantaneous spin configuration at every time step. Repeated quantum calculations would be prohibitively expensive for large-scale simulations. Consequently, empirical classical spin Hamiltonians, from which the local fields can be explicitly calculated, are often employed in large-scale dynamical simulations of skyrmion magnets~\cite{Evans_2014,mumax}. Yet, such classical spin models often cannot capture the intricate long-range spin-spin interactions mediated by electrons.

The computational complexity of the above quantum approaches to spin dynamics is similar to the so-called quantum or {\em ab~initio} molecular dynamics (MD) methods. Contrary to classical MD methods that are based on empirical force fields, the atomic forces in quantum MD are computed by integrating out electrons on-the-fly as the atomic trajectories are generated~\cite{marx09}. Various many-body methods, notably the density functional theory, have been used for the force calculation of quantum MD. However, the computational cost of repeated electronic structure solutions significantly restricts the accessible scales of atomic simulations. To overcome this computational difficulty, machine learning (ML) methods have been exploited to develop force-field models by accurately emulating the time-consuming many-body calculations, thus enabling large-scale MD simulations with the desired quantum accuracy. 

Crucial to the remarkable scalability of ML-based force-field models is the divide-and-conquer approach proposed in the pioneering works of Behler and Parrinello~\cite{behler07}, and Bart\'ok {\em et al.}~\cite{bartok10}. In this approach, the total energy of the system is partitioned into local contributions $E = \sum_i \epsilon_i$, where $\epsilon_i$ is called the atomic energy and only depends on the local environment of the $i$-th atom~\cite{behler07,bartok10}. The atomic forces are then obtained from the derivatives of the predicted energy:~$\mathbf F_i = -\partial E / \partial \mathbf r_i$, where $\mathbf r_i$ is the atomic position vector. Crucially, the complicated dependence of atomic energy $\epsilon_i$ on its local neighborhood is approximated by the ML model, which is trained on the condition that both the predicted individual forces $\mathbf F_i$ as well as the total energy $E$ agree with the quantum calculations~\cite{behler07,bartok10,li15,botu17,li17,smith17,zhang18,behler16,deringer19,mcgibbon17,suwa19,mueller20}. It is worth noting that physically the principle of locality, or the so-called nearsightedness of electronic matters, lies at the heart of this approach~\cite{kohn96,prodan05}. 

The tremendous success of ML methods in quantum MD simulations has spurred similar approaches to multi-scale dynamical modeling of other functional electronic systems in condensed matter physics~\cite{zhang22a,ma19,liu22,zhang22b,zhang20,zhang21,zhang23}. In particular, the Behler-Parrinello (BP) ML scheme~\cite{behler07,bartok10} was generalized to build effective magnetic energy or torque-field models with the accuracy of quantum calculations for itinerant electron magnets~\cite{zhang20,zhang21,brannvall22,novikov22}. Notably, large-scale dynamical simulations enabled by such ML models uncovered intriguing phase separation dynamics that results from the nontrivial interplay between electrons and local spins. While the conventional BP scheme can only represent conservative forces, a generalized potential theory for the Landau-Lifshitz equation allows one to extend the BP scheme to describe non-conserved spin torques that are crucial to the dynamical modeling of out-of-equilibrium itinerant spin systems~\cite{zhang23}.

In this paper, we present an ML torque model for itinerant magnets based on convolutional neural networks (CNN). CNN is a class of neural networks that can be characterized by its local connectivity, implemented via finite-sized convolution kernels. Importantly, the convolution operation with a finite-sized kernel naturally incorporates the locality principle into the ML structure, thus offering an efficient implementation of the ML torque model that can be straightforwardly scaled to larger systems. Our CNN model is designed to directly predict the vector torque field at every site without the need for the introduction of local energies as in the BP scheme. Data augmentation techniques are employed to incorporate the spin-rotational symmetry and the lattice symmetry into the CNN spin-torque model. We demonstrate our approach on an itinerant spin model which exhibits a skyrmion crystal phase at an intermediate magnetic field. We show that dynamical simulations with magnetic torques computed from the trained CNN model faithfully reproduce the relaxation process of the itinerant spin systems. Moreover, the CNN model, while trained by datasets from small systems, is capable of stabilizing a skyrmion lattice on larger systems, thus demonstrating the transferability and scalability of our ML approach.

The rest of the paper is organized as follows. In Section~\ref{sec:mag-dyn}, we discuss the methods for simulating the spin dynamics of itinerant electron magnets. A triangular-lattice s-d model, a well-studied itinerant spin system, is used as a concrete example to highlight the complexity of the dynamical simulations. We also briefly review BP-type ML approaches, where a neural network is trained to approximate a local energy function. Section~\ref{sec:method} presents the CNN structure used for the prediction of spin-torque. Details of the data augmentation for incorporating symmetries and how the ML model can be scaled to larger systems are also discussed. Using the s-d model as an example, a benchmark of the CNN models and simulation results based on the trained models are presented in Section~\ref{sec:results}. We also ascertain the scalability and symmetry of the proposed CNN method, as well as its compliance with the locality principle. Finally, we summarize our work and discuss future directions in Sec.~\ref{sec:conclusion}.

\section{magnetization dynamics of the itinerant magnets}

\label{sec:mag-dyn}

The magnetization dynamics in spin systems is governed by the Landau-Lifshitz-Gilbert (LLG) equation~\cite{brown63}
\begin{eqnarray}
	\label{eq:LLG}
	\frac{d \mathbf S_i}{dt} = \mathbf T_i  - \alpha \,\mathbf S_i \times \mathbf T_i  + \bm \tau_i ,
\end{eqnarray}
where $\mathbf T_i$ is the magnetic torque defined as
\begin{eqnarray}
	\mathbf T_i = \gamma \mathbf S_i \times \mathbf H_i.
\end{eqnarray}
Here $\gamma$ is the gyromagnetic ratio, and $\mathbf H_i$ is an effective exchange field acting on spin-$i$, $\alpha$ is the damping coefficient, and $\bm\tau_i(t) = \mathbf S_i \times \bm\eta_i(t)$ is a fluctuating torque generated by a random local field $\bm\eta_i$ of zero mean. The stochastic field~$\bm\eta_i$ is assumed to be a Gaussian random variable with the variance determined from $\alpha$ and temperature $T$ through the fluctuation-dissipation theorem. The LLG simulations are widely used to study dynamical phenomena in a wide range of magnetic systems, including spin waves in unusual magnetic phases and dynamical behaviors of skyrmions and other spin-textures. 

For adiabatic spin dynamics, the local exchange field is given by the derivative of the system energy $E = E(\mathbf S_i )$:
\begin{eqnarray}
	\label{eq:H-force}
	\mathbf H_i = - \frac{\partial E}{\partial \mathbf S_i}.
\end{eqnarray}
For magnetic insulators, interactions between spins are often short-ranged. The resultant magnetic energy has the form of bilinear interactions between a few nearest-neighbor spins on the lattice, e.g. $E = \sum_{ij} (J_{ij} \mathbf S_i \cdot \mathbf S_j + \mathbf D_{ij} \cdot \mathbf S_i \times \mathbf S_j)$, where $J_{ij}$ denotes the isotropic Heisenberg exchange interaction and $\mathbf D_{ij}$ represents the anisotropic exchange, also known as the Dzyaloshinskii-Moriya interaction~\cite{Evans_2014,mumax}. The exchange field of such models is explicitly given by $\mathbf H_i = - \sum_j (J_{ij}  \mathbf S_j + \mathbf D_{ij} \times \mathbf S_j)$, where the summation is restricted to a few nearest neighbors, and can be very efficiently computed for large-scale LLG simulations.

On the other hand, the exchange fields in a metallic magnet come from interactions between local spins and itinerant electrons. Here we consider spin dynamics in the adiabatic approximation, which is analogous to the Born-Oppenheimer approximation in quantum molecular dynamics~\cite{marx09}. In the adiabatic limit, electron relaxation is assumed to be much faster than the time scale of local magnetic moments. As a result, the magnetic energy $E$ in Eq.~(\ref{eq:H-force}) can be obtained by freezing the spin configuration and integrating out the electrons. The resultant spin-dependent energy function, $E = E(\mathbf S_i )$, can be viewed as a potential energy surface (PES) in the high-dimensional spin space, similar to the PES in Born-Oppenheimer MD simulations. In practice, the calculation of this magnetic PES requires solving the electron structure problem that depends on the instantaneous spin structure $\{\mathbf S_i(t)\}$. 

For concreteness, here we consider a generic single-band s-d model for such itinerant magnets. The s-d model describes the interaction between itinerant $s$-band electrons and magnetic moments $\mathbf S_i$ of localized $d$-electrons. Its Hamiltonian reads
\begin{eqnarray}
	\label{eq:H_sd}
	\mathcal{H} = \sum_{ ij}\sum_{\alpha = \uparrow,\downarrow} t_{ij} {c}^{\dagger}_{i \alpha} {c}^{\;}_{j \alpha} 
	- J \sum_{i}\sum_{\alpha,\beta = \uparrow,\downarrow} \mathbf S_i \cdot c^{\dagger}_{i\alpha}  {\bm{\sigma}_{\alpha\beta}} c^{\;}_{i\beta}, \quad
\end{eqnarray}
where ${c}^\dagger_{i \alpha}/c_{i, \alpha}$ are creation/annihilation operators of an electron with spin $\alpha = \uparrow, \downarrow$ at site $i$, $t_{ij}$ is the electron hopping constant between a pair of sites $(i, j)$, $J$ denotes the strength of local Hund's rule coupling between electron spin and magnetic moment $\mathbf S_i$ of localized $d$-electrons. For most skyrmion magnets, these local magnetic moments can be well approximated as classical spins of fixed length $|\mathbf S_i | = S$. 

For small Hund's coupling $J \ll t_{ij}$, the effective energy of spins can be obtained by integrating out electrons via a second-order perturbation calculation, giving rise to a long-ranged oscillatory interaction, similar to the so-called Ruderman-Kittel-Kasuya-Yosida (RKKY) interaction~\cite{ruderman54,kasuya56,yosida57}. However, for intermediate and large Hund's coupling,  the effective energy to be used for the force calculation in Eq.~(\ref{eq:H-force}) has to be obtained by integrating out electrons on the fly:
\begin{eqnarray}
	\label{eq:E_system}
	E = \langle \mathcal{H} \rangle = {\rm Tr}(\rho \mathcal{H}),
\end{eqnarray}
where $\rho = \exp(-\mathcal{H}/k_B T)$ is the density matrix of the equilibrium electron liquid within the adiabatic approximation. The calculation of the density matrix, in the absence of electron-electron interaction, amounts to solving a disordered tight-binding Hamiltonian for a given spin configuration. The standard method for solving tight-binding models is based on exact diagonalization, whose complexity scales cubically with the system size. As a result, for large-scale LLG simulations of the s-d model, repeated ED calculations of the electron density matrix can be overwhelmingly time-consuming. 

As discussed in Sec.~\ref{sec:intro}, the BP scheme has been generalized to develop ML-based models for the effective spin energy $E(\{\mathbf S_i\})$ of itinerant magnets~\cite{zhang20,zhang21,brannvall22,novikov22,zhang23}. In this approach, the total energy is partitioned into local contributions 
\begin{eqnarray}
	E = \sum_i \epsilon_i =\sum_i \varepsilon(\mathcal{C}_i),
\end{eqnarray}
where the energy $\epsilon_i = \varepsilon(\mathcal{C}_i)$ is associated with the $i$-th lattice site and is assumed to depend only on spin configuration $\mathcal{C}_i = \{ \mathbf S_j \mid \|\mathbf r_j - \mathbf r_i\| < r_c \}$ in its neighborhood. This local energy function $\varepsilon(\mathcal{C}_i)$ can be viewed as the building block of the magnetic PES.  Importantly, the complicated dependence of the PES on the neighborhood spins is to be approximated by fully connected neural networks~\cite{zhang20,zhang21,zhang23}.  To preserve the $\mathrm{SO} (3)$ spin rotation symmetry, the inner product between spin pairs $b_{jk} = \mathbf S_j \cdot \mathbf S_k$ and scalar product between spin triplets $\chi_{jkl} = \mathbf S_j \cdot\mathbf S_k \times \mathbf S_l$ within the neighborhood are used as building blocks to construct feature variables that are input to the neural network. Finally, exchange fields $\mathbf H_i$ acting on spins are obtained by applying automatic differentiation to the ML energy model.

\section{\label{sec:method}CNN spin torque model}

\begin{figure*}
\centering
    \includegraphics[width=\textwidth]{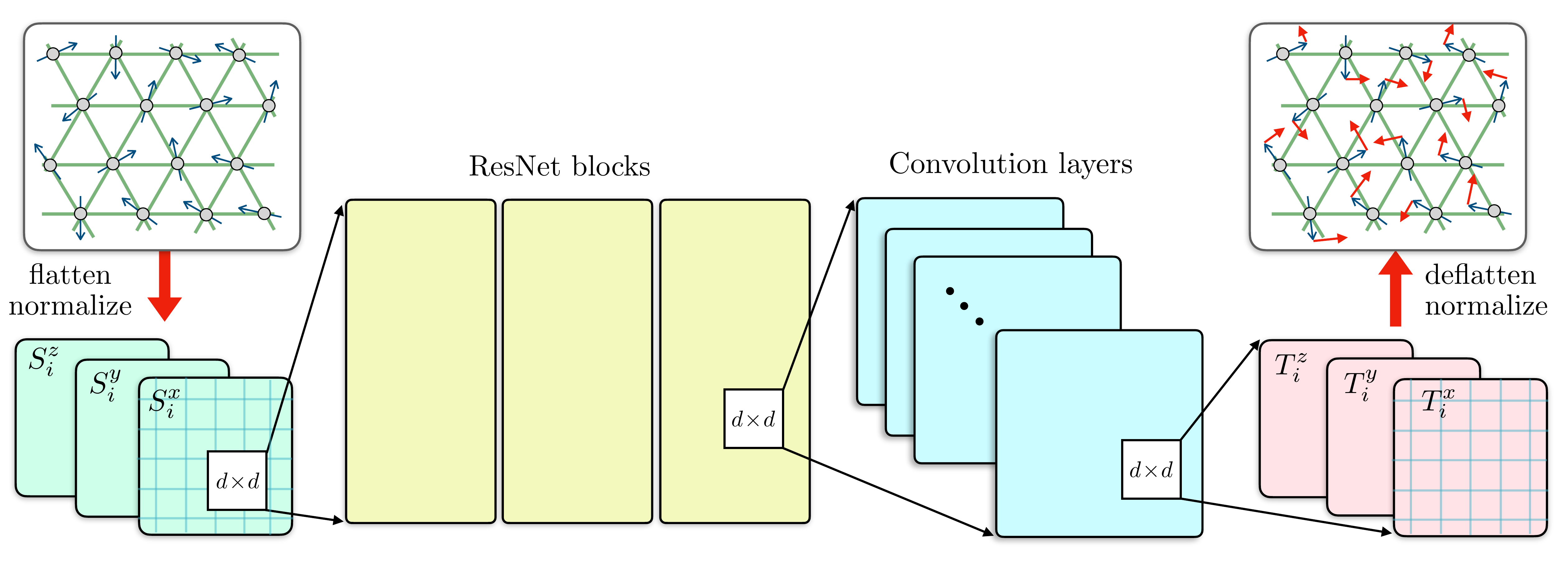}
    \caption{Schematic diagram of the CNN-based ML model for spin-torque  prediction of itinerant electron magnets. The spin configuration $\{\mathbf S_i\}$ on a lattice is first flattened to give three arrays, corresponding to the three components of spins, which are input to a series of ResNet blocks. Details of the ResNet are presented in Fig.~\ref{fig:resnetblock_diagram}. The output of the ResNet blocks is then processed through additional convolution layers. The final output are three arrays which after deflattening correspond to the torques $\{\mathbf T_i\}$ that drive the spin dynamics. }
    \label{fig:ml_diagram}
\end{figure*}

The BP-type schemes described here essentially provide an energy-based ML model for force field calculations. A crucial step is the partitioning of the total energy into local contributions $\epsilon_i$, which cannot be directly computed from electronic structure methods that are used to generate the training dataset. As a result, the loss function $L$ cannot be directly determined from the predicted energies $\epsilon_i$. Instead, it is constructed from the, \textit{e.g.,} mean square error (MSE) or ``forces", or in our case, the spin torque fields, and only implicitly depends on the predicted energy through automatic differentiation. However, the uncertainties due to the introduction of such intermediate local energies often complicate the training of BP-type models. While one advantage of the BP-type scheme is the explicit inclusion of the physical constraint of conservative forces, such energy-based ML approaches, however, are also restricted to the representation of only conservative forces. In this Section, we present an alternative ML approach that directly predicts the vector forces without going through intermediate energy.

\subsection{Convolutional Neural Networks}
\label{sec:convnet}

The fact that spins in metallic magnets are defined on well-known lattices suggests that spin configurations can be treated as generalized ``images," which can then be processed using powerful image-processing techniques developed in recent years, such as CNNs.
Below, we present a CNN model for the direct prediction of torques $\mathbf T_i$ that drive the spin dynamics. As illustrated in Figure~\ref{fig:ml_diagram}, the proposed network takes the spin configuration $\{\mathbf S_i\}$ on the lattice as input and returns the torques $\{\mathbf T_i\}$ that drive the spin dynamics as output. The model is comprised of multiple convolution layers $f_m$ with associated activation (nonlinearity) layers $\sigma_m$ to model the complex nonlinear relationship between $\{\mathbf S_i\}$ and $\{\mathbf T_i\}$ as a composition of such layers: $f_\text{CNN} = (\sigma_L \circ f_L) \circ \cdots \circ (\sigma_1 \circ f_1)$, where $L$ is the number of layers or the \textit{depth} of the CNN model.

Given an input vector field $V \in C^\infty (\R^2, \R^d)$, each convolution layer $f_m$ maps the vector field onto an output vector field  $W \in C^\infty (\R^2, \R^k)$ by convolving a \textit{kernel} tensor field $h_m(X):=h(X; \theta_m)$ with trainable parameters $\theta_m$, via the convolution operation: 
\begin{eqnarray}
    W(\mathbf{r}) := \int_{\R^2}V(\mathbf{q})h_m(\mathbf{r}-\mathbf{q})d\mathbf{q}.
    \label{eq:convolution}
\end{eqnarray} 
Each vector element of the vector field $W$ then undergoes the activation function $\sigma_m:\R \rightarrow \R$ to produce the output vector field $A \in C^\infty (\R^2, \R^k)$, called \textit{activation maps}. A variety of activation functions can be used in CNNs. In the current work, we use the \textit{rectified linear unit}, or \textit{ReLU} \citep{nair2010} as an activation function:
\begin{eqnarray}
    \sigma_m(x) := \max (0, x).
    \label{eq:relu}
\end{eqnarray}
for $m=1,\hdots,L-1$. Note that the final layer $f_L$ has no activation function associated with it, or technically, $\sigma_L(x) = x$.

Typically in CNNs, the support of a kernel $\supp(h_m)$, \textit{i.e.,} the region where $h_m$ has nonzero values (also known as the \textit{receptive field} of the kernel), is limited to a small region (\textit{e.g.,} $5\times 5$ lattice sites) such that the activation response $W(\mathbf{r})$, thereby $A(\mathbf{r})$, at position $\mathbf{r}$ is limited to the patterns of $V$ only within the close proximity of $\mathbf{r}$. It is worth noting that the physical justification of employing such finite-size kernels is the principle of locality: \textit{viz.,} local physical quantities, such as local spin torque $\mathbf T_i = \mathbf T(\mathbf r_i)$, are predominately determined by the local environment of site-$i$:
\begin{eqnarray}
	\mathbf T_i = \boldsymbol{\mathcal{T}}(\mathcal{C}_i),
\end{eqnarray} 
where $\mathcal{C}_i$ is the magnetic environment in the vincinity of site-$i$, and the vector function $\boldsymbol{\mathcal{T}}(\cdot)$ is to be modeled by the CNN. The range of the neighborhood $\mathcal{C}_i$ is determined by the sizes of kernels and the number of convolution layers.

Note that Eqs.~(\ref{eq:convolution})~\&~(\ref{eq:relu}) imply that the output activation $A(\mathbf{r})$ at position $\mathbf{r}$ will be of a large, positive magnitude, only if the input vector field $V(\mathbf{r})$ is closely correlated to the (shifted) kernel $h_m(\mathbf{q} - \mathbf{r})$. Therefore, the goal of training the CNN is to find the unknown kernel parameters $\theta_{m=1,\hdots,L}$ that determines the function shapes of $\{h_m\}$, in order to produce the adequate activation values such that the final output of the model $f_\text{CNN}(\{\mathbf S_i\})$ can reasonably approximate the ground truth spin torque $\{\mathbf T_i\}$ in the training data.

Meanwhile, the composition of convolution layers enables hierarchical modeling of the spin-torque relationship. That is, while an individual kernel limited to a small region may only represent rather simplistic patterns (\textit{e.g.,} small blobs), the composition of such kernels across layers alongside the nonlinear ReLU activation can produce fairly complex, nonlinear patterns. Furthermore, the composition of convolution layers $(\sigma_n \circ f_n) \circ (\sigma_m \circ f_m)$ in effect produces a larger receptive field area, equal to the Minkowski sum\footnote{Note that the definition of receptive fields using the notion of Minkowski sum may only be applicable to our particular setting, in which we are considering both the input and output vector fields over the same domain $\R^2$.} of the receptive fields of the individual layers $\text{supp}(h_n \otimes h_m) = \text{supp}(h_n) \oplus \text{supp}(h_m)$. Therefore, stacked convolution layers produce a natural hierarchy, in which earlier layers represent local, primitive patterns in a small proximity while latter (deeper) layers represent more global, sophisticated patterns in a relatively larger periphery.

Finally, a purely convolutional CNN, without any conventional fully connected (dense) layers, can restrict the overall receptive field size of the entire model $\text{supp}(h_L \otimes \cdots \otimes h_1)$ to a predetermined lattice size, presenting a distinct advantage of \textit{built-in locality}. 
Further, with a purely convolutional design, since the sizes of kernels in a CNN are fixed for a given itinerant model, the successfully trained CNN model can then be used in much larger lattice systems without the need to rebuild or retrain a new neural network. The CNN structure thus provides a natural approach to implementing scalable ML models based on the locality principle.

\subsection{Model Architecture}
\label{sec:architecture}

%(\textbf{[Phong: Xinlun, you did not use ResNet directly but design one similar to it right, if so, I think we better say our design is inspired by ResNet, otherwise, we may lose some credit]})

%Our model directly predicts the exchange field component perpendicular to the spin of $i$-th lattice site. In doing so, the model does not rely on the calculation of the system's energy first, in order to allow the calculation of non-conservative forces as the work of a nonconservative force depends on the path taken and has no associated potential energy with it. %({\color{red}this needs further elaboration}).

%\textbf{[Phong: I think we need a more formal way of describing input and output of the network. For instance, the network take input tensor with shape BxHxWxC and return output tensor of shape BxHxWxC. In the input tensor, xxx represent the spin vector ... Meanwhile, the output tensor, xxx represent x,y,z component of the forces... I am making this up. I think it is better for reader to understand what are we predicting and from what.]}

%In other words, the output response at the $i$-th lattice site only depends on the input values within the local neighborhood of $i$-th lattice site. Such a locality of CNN is the key to modeling the near-sightedness of electronic matters.

%and built-in \textit{scalability}. 

\Cref{fig:ml_diagram} shows a schematic diagram of our CNN architecture. The input to the CNN is the spin vector field $\{\mathbf S_i\}$ transformed from triangular lattice to square (``flattening'') as most convolution operation expects a square input.
%the three components of spins on the lattice. ({\color{red}flattend?}). 
We employed four ResNet blocks, inspired by He \textit{et al.}~\citep{he2015}, as our backbone, which processes the input into an activation map of 512 features. These features then undergo two additional convolution layers, which in the end produce the torques as the output of the network. In our model, these torque vector outputs are obtained in a normalized range of values with a mean magnitude of 1. Such normalized outputs are then scaled by the mean magnitude of the torque vectors in the training data set. Finally, the predicted torques on the square grid are ``unflattened'' onto the original triangular lattice.

Compared to fully connected (dense) layers, in which each neuron aggregates values across the entire domain into a single scalar value via the weighted sum, convolution layers preserve the spatial structure of the input domain. Therefore, with the proper boundary condition (\textit{e.g.}, `padding' in the machine learning jargon), the output lattice is guaranteed to have the same size and resolution as the input lattice. Therefore, a CNN comprised of purely convolution layers, without any fully connected layers, can be scaled to an arbitrary lattice of practically any size, as long as the lattice element has locally the same geometric and topological structures.

\begin{figure}[!ht]
    \centering
    \includegraphics[width=0.8\columnwidth]{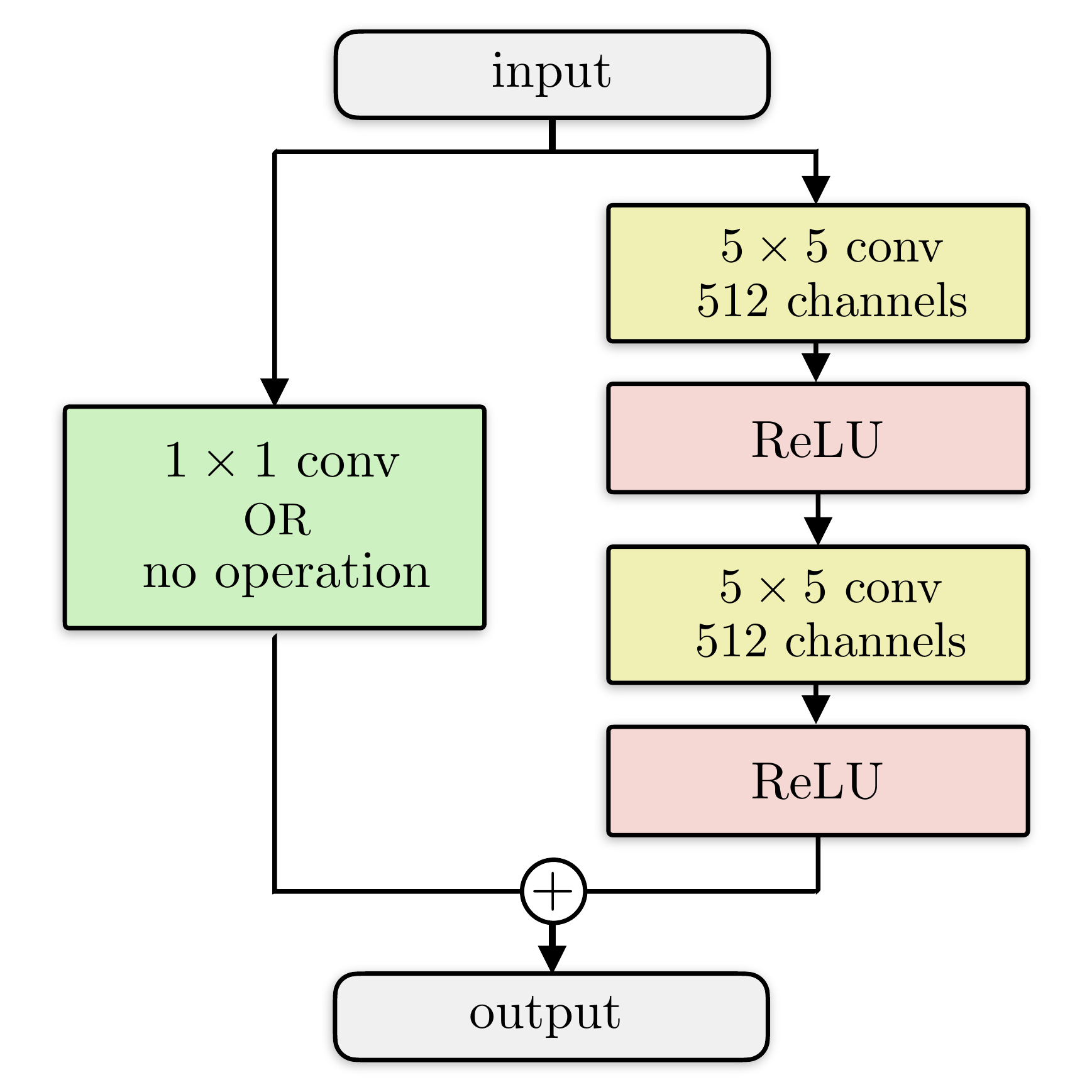}
    \caption{Diagram of a ResNet block. The input to a Resnet block goes through two different pathways: the skip connection, where no operation is performed if input and output have the same number of channels or a $1 \times 1$ convolution layer otherwise; and normal connection, where two $5\times 5$ convolution - ReLU activation blocks are stacked on top of each other. The results of the two pathways are then added together as the output.}
    \label{fig:resnetblock_diagram}
\end{figure}

Meanwhile, the architecture of the ResNet block in the backbone is described in \Cref{fig:resnetblock_diagram}. Similar to the original ResNet, the input goes through two separate pathways. One pathway (\textit{right} path in \Cref{fig:resnetblock_diagram}) is comprised of two convolution layers with the ReLU activation \citep{nair2010}, stacked on top of each other, in order to develop a feature vector characterizing patterns of the input vector field at each local neighborhood on the pixel grid. The other pathway (\textit{left} path in \Cref{fig:resnetblock_diagram}) can be either the skip connection, in which the input values are directly copied without any transformation, or a 1$\times$1 convolution, in which the input feature vector at each grid location undergoes a dimensionality reduction. The outputs from the two pathways are then added together to produce the overall output of the ResNet block. Here, note that we do not employ the batch normalization technique, which is a technique used in the original ResNet model to avoid the vanishing gradient problem. Empirically, we found that batch normalization overly regularized the network causing severe underfitting of the spin torque and overall deteriorating the prediction performance. Moreover, since the input spin vectors are already well normalized to have a length of 1, it is not necessary to employ batch normalization.

%In our experimentation, we found that the batch normalization layers in the original ResNet paper deteriorate the performance of our model, causing underfitting in the Z component of the exchange field. Hence, no batch normalization was adopted in our design. We would also point out that since the magnitude of the spin vector is guaranteed to be 1 and the field has been normalized to have a length of 1, we do not believe that further normalization is necessary in the middle of the model 
%({\color{red}I don't know what y}) \textbf{[XC: I think this is one of Phong's comments when he suggested ways to fix the underfitting/bias at higher force values.]}. \textbf{[Phong: I revised the text to make this point clearer. Please check.]}

%range of values {\color{red}$[XXX, YYY]$}. From this, the final local exchange field prediction can be calculated by multiplying the output by a constant (obtained during training) and reshape back into the original triangular lattice shape ({\color{red}this sentence needs further elaboration}).

\subsection{\label{subsec:training}Training} 

%Our training and testing set consist of a combination of simulation runs using analytical local exchange field calculation of two different types of initial condition on a 48$\times$48 triangular lattice: perturbed Skyrmion, where the periodic Skyrmion crystal structure was baked into the initial condition but spins had random noise added, and random initial condition, where the initial spins were randomly generated ({\color{red}This sentence is too long and very difficult to comprehend. Please break it to smaller sentences}). 

Our training and testing sets consist of 60 independent spin dynamics simulations, respectively, performed on a 48$\times$48 triangular lattice. The following parameters are used for the s-d Hamiltonian Eq.~(\ref{eq:H_sd}): The nearest-neighbor hopping was set to $t_1 = 1$, which also provides the reference unit for energy. A third-neighbor hopping $t_3 = -0.85$ is included in order to stabilize a triple-$Q$ magnetic order that underlies the skyrmion lattice (SkL) phase~\cite{ozawa17}. The electron-spin coupling constant is set at $J = 1$. An electron chemical potential $\mu = -3.5$ was used, and an external magnetic field $H_{\rm ext} = 0.005$ was included to explicitly break the time-reversal symmetry and induce the SkL~\cite{ozawa17}. As discussed in Section~\ref{sec:mag-dyn},  the exchange fields $\mathbf H_i$ acting on spins are obtained by solving the electron Hamiltonian. Specifically, using Eq.~(\ref{eq:H-force}) and the s-d Hamiltonian Eq.~(\ref{eq:H_sd}), the exchange fields are given by 
\begin{eqnarray}
	\mathbf H_i = J \sum_{\alpha,\beta =\uparrow, \downarrow} \bm\sigma^{\,}_{\alpha\beta} \, \rho^{\,}_{i\beta, i \alpha},
\end{eqnarray}
where $\rho_{i\alpha, j\beta} := \langle c^\dagger_{j\beta} c^{\,}_{i\alpha} \rangle$ is the electron correlation function, or single-electron density matrix. The kernel polynomial method (KPM)~\cite{weisse06,wang18} was used to compute the electron density matrix for generating the training dataset. The KPM is more efficient compared with exact diagonalization, yet is considered numerically exact when a large number of Chebyshev polynomials and random vectors are used.

The time-scale of the precession dynamics of the LLG equation~(\ref{eq:LLG}) is given by $t_0 = (\gamma J S)^{-1}$, where $\gamma$ is the gyromagnetic ratio, $J$ is the electron-spin coupling, and $S$ is the length of the localized magnetic moments. The damping term introduces another time-scale $t_{\rm damping} = t_0 /\alpha$ which characterizes the rate of energy dissipation, where $\alpha$ is a dimensionless coefficient. In the following, the simulation time is measured in terms of $t_0$, and a damping coefficient $\alpha = 0.05$ is used. 

The initial conditions of the simulations are divided into two types. The first one is perturbed SkL, where a periodic array of skyrmions  is baked into the initial condition but spins had random noise added. The other type is random initialization where the spins are totally randomly generated. For each type of initial condition, a total of 30 simulations were generated, each of them comprised of 5,000 time steps. %(equivalent to 5000 seconds). 
For a given initial condition, a semi-implicit second-order scheme~\cite{mentink10} which preserves the spin length was employed to integrate the LLG equation~(\ref{eq:LLG}) with a time-step $\Delta t = 0.1$. 

\begin{figure*}
    \centering
    \includegraphics[width=0.96\textwidth]{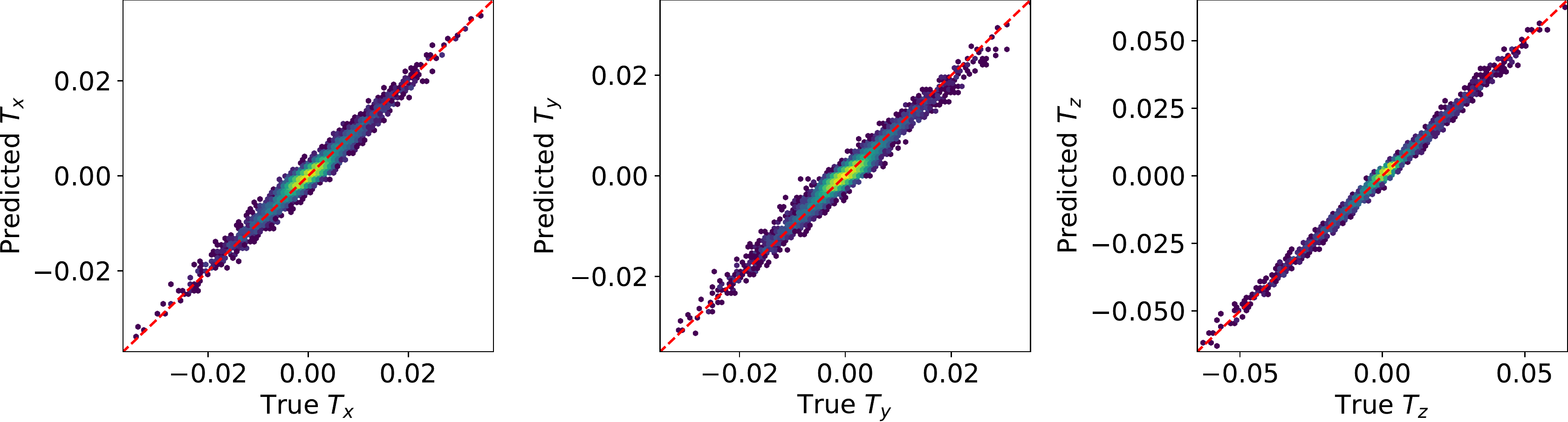}
    \includegraphics[width=0.96\textwidth]{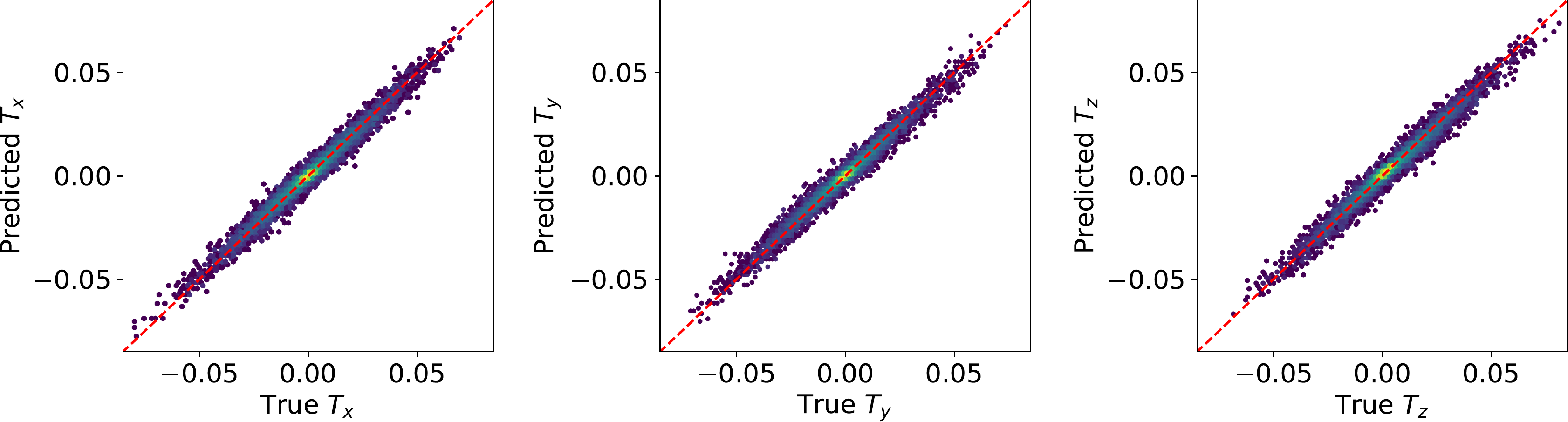}
    \caption{Predicted spin torque components $(T_x, T_y, T_z)$ versus ground truth components from the testing set. The red-dotted diagonal lines indicate perfect prediction. The top row shows the prediction results based on spin configurations obtained from LLG simulations of a perturbed SkL. Results from LLG simulations with random initial states are shown in the bottom row.}
    \label{fig:pred_true_xyz}
\end{figure*}

The spins and their corresponding exchange fields at all lattice sites were collected every 10 other steps in the simulation. We focused on the training of the electron-induced exchange field, so the external constant field of $H_{\rm ext} = 0.005$ in the $z$ direction was  removed. The field $\mathbf H_i$ was then decomposed into components that are parallel and perpendicular to spin components, and only the perpendicular component, which is equivalent to the torque $\mathbf T_i$, was kept as the parallel component has no effect in the evolution of spin configuration and is around two orders of magnitude larger than the perpendicular component. The perpendicular fields were then normalized to have a mean magnitude of 1 over the entire dataset. Then 70\% of the entire dataset was used for the training, while the rest was set aside for validation.  The split of the dataset is stratified so the training and testing set has the same proportion of the two types of simulations.

%The local energy function $\varepsilon(\mathcal{C}_i)$, which can be viewed as an effective energy after integrating out electrons, should retain the global SO(3) rotation symmetry of classical spins $\mathbf S_i$. 

The triangular-lattice s-d Hamiltonian in Eq.~(\ref{eq:H_sd}) is invariant under two independent symmetry groups: the $\mathrm{SO} (3)$/$\mathrm{SU} (2)$ rotation of spins and the $\mathrm{D}_6$ point group of the triangular lattice. Here the rotation symmetry refers to the global rotation of local magnetic moments $\mathbf S_i \to \mathcal{R}\cdot \mathbf S_i$ (treated as classical vectors), and a simultaneous unitary transformation of the electron spinor $\hat{c}_{i\alpha} \to \hat{U}_{\alpha\beta} \hat{c}_{i\beta}$, where $\mathcal{R}$ is an orthogonal $3\times 3$ matrix and $\hat{U} = \hat{U}(\mathcal{R})$ is the corresponding $2\times 2$ unitary rotation operator. The ML model, corresponding to an effective force-field model by integrating out electrons, needs to preserve the $\mathrm{SO} (3)$ rotation symmetry of spin, which means under uniform rotation $\mathcal{R}$ of all spins in the neighborhood, the ML predicted spin torques should undergo the same rotation transformation $\mathbf T_i \to \mathcal{R}\cdot \mathbf T_i$. 
On the other hand, under a symmetry operation $g$ of the $\mathrm{D}_6$ point group centered at some lattice point, both spins and torques transform under the $\mathrm{D}_6$ point group as: $\mathbf S_i \to \mathbf S_j$ and $\mathbf T_i \to \mathbf T_j$, where the lattice points $\mathbf r_j = \mathcal{R}(g) \cdot \mathbf r_i$, and $\mathcal{R}(g)$ denotes the $3\times 3$ matrix corresponding to $g$.

To incorporate both symmetries into the CNN model, we introduced data augmentation during our training phase. Specifically, for each input spin configuration and the corresponding torque field, a random $\mathrm{SO} (3)$ rotation field was applied to the spins $\{\mathbf S_i\}$ and a random $\mathrm{D}_6$ symmetry operation was applied to the lattice points. The same symmetry operations, both for spin-space and real-space lattice, were also applied to the torque fields $\{\mathbf T_i\}$. These additional symmetry-generated input/output configurations were included along with the original ones to the dataset for supervised training. We note that, contrary to previous ML models where the symmetry is explicitly included through descriptors, the symmetry of the itinerant electron Hamiltonian is enforced on the ML vector model statistically in this data-driven approach. 

Since the torques in the dataset could differ at most by one order of magnitude, we find that the usual mean absolute error or mean square error loss functions do not perform very well. Instead, we adopted a mean percentage absolute loss:
\begin{eqnarray}
\label{eq:loss}
L = \frac{1}{N} \sum_{i=1}^N \frac{|T^x_{i}-\hat{T}^x_{i}|+|T^y_{i}-\hat{T}^y_{i}|+|T^z_{i}-\hat{T}^z_{i}|}{|\hat{\mathbf T}_i|}, \,\,
\end{eqnarray}
where $N$ is the total number of lattice sites within each batch summed across all lattices, $ \hat{\mathbf T}_{i}$ is the ground truth field vector at $i$-th lattice site and $\mathbf T_{i}=(T^x_{i}, T^y_{i}, T^z_{i})$ is the predicted field vector and its three components.

An Adam \citep{kingma2017} optimizer with an initial learning rate of $10^{-3}$ was used for the training. The learning rate was later reduced to $10^{-6}$ upon the plateau of loss value in the testing set. We did not use any regularization methods, such as dropout or weight decay, and there is no evidence of overfitting when comparing training and testing set loss values. The model and its training process are implemented in PyTorch \citep{adam2019}, and training was performed on one Nvidia A100 for roughly 72 hours. %\textcolor{blue}{[cite github?]} %The trained model can be accessed at ...

\section{\label{sec:results}Results}

Here we present benchmarks of the CNN models by comparing spin torque predictions and small-scale dynamical simulations against exact methods.  We further demonstrate the restoration and stability of a skyrmion lattice in large-scale LLG simulations, highlighting the scalability and transferability of our ML approach.

\subsection{Benchmark of Spin Torque Prediction}

The spin torques $\mathbf T_i$ predicted from the trained CNN model are compared against the ground truth in \Cref{fig:pred_true_xyz} using configurations from the test dataset. Two types of testing data are used for this benchmark: LLG simulations of an initially perturbed SkL state, and LLG simulations starting from random spins. In both cases, the predicted torque components closely follow the ground truth with roughly equal variance across the entire range. Note the values of torque components in the random spin case span a range nearly twice larger than that of the SkL case. As can be expected, the ML model performs better in the case of the SkL simulations since spin configurations here correspond to a rather small and special set of the whole state space. Yet, a fairly good agreement was obtained even for the testing dataset with completely random initial spins.

%The most straight forward way to characterize the performance of the machine learning model is to directly compare the predicted local exchange field to the ground truth. In \Cref{fig:pred_true_xyz}, we present such a comparison of field components (x, y, z) to those from ground truth as seen in our testing set. As can be clearly seen, regardless of input spins, the field predictions closely follow the ground truth with roughly equal variance across the entire value range. Overall the machine learning model performs better when the input is perturbed Skyrmion spin configuration. This is expected as field calculation with random spin configuration is substantially more complex than perturbed Skyrmion.

\begin{figure}[t]
    \centering
    \includegraphics[width=\columnwidth]{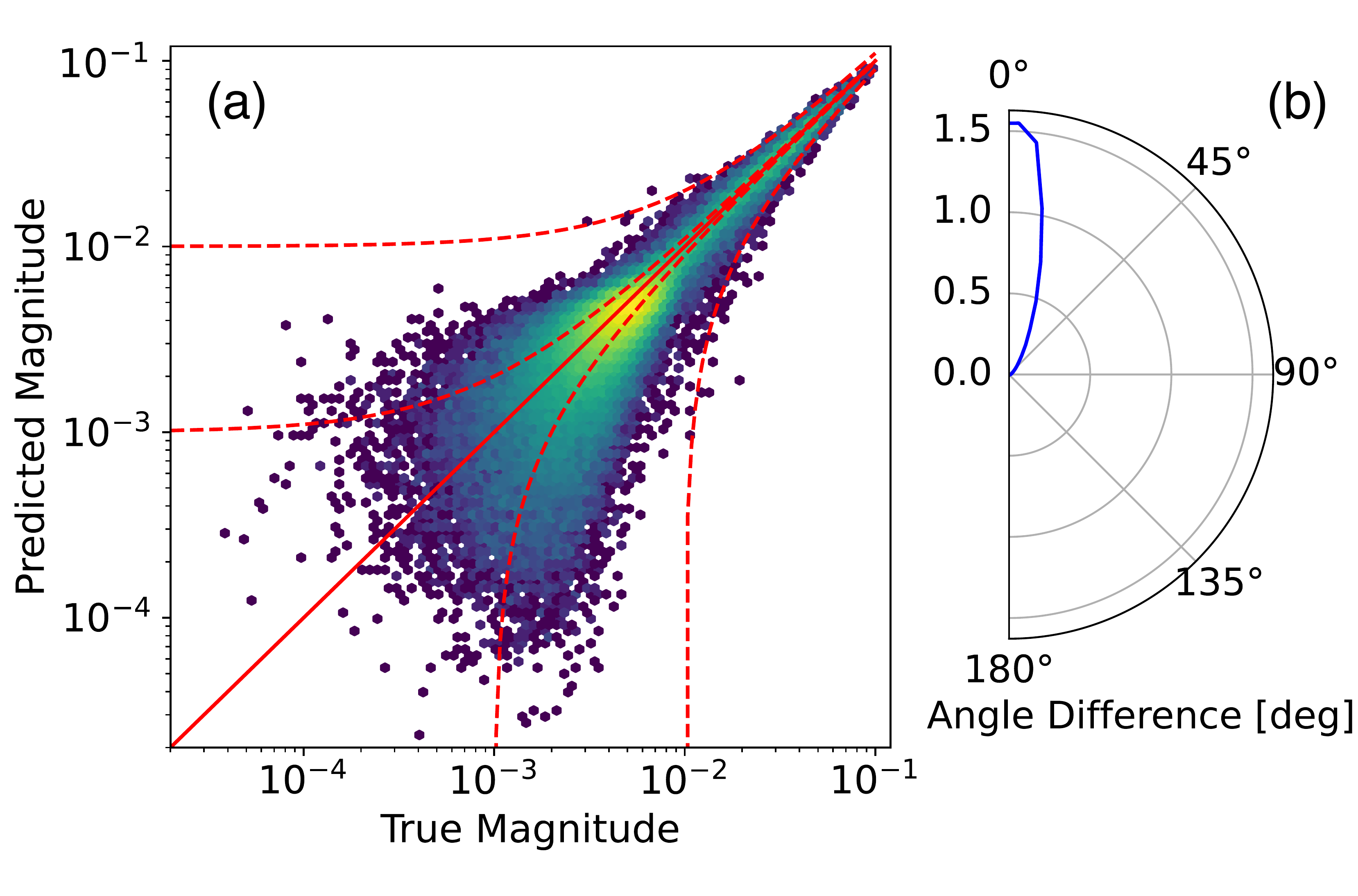}
    \caption{(a) Comparison of predicted field vector magnitude against ground truth. The red line indicates prediction equalling to ground truth, the outer red dotted line represents a $10^{-2}$ deviation from ground truth magnitude while the inner one represents a deviation of $10^{-3}$. The color denotes the log density. (b)  Angular difference between ground truth field vector and predicted field vector.}
    \label{fig:pred_true_mag}
\end{figure}

We further examine the magnitude of predicted torques versus the ground truth, as well as the angle between the predicted field vector and ground truth vector in \Cref{fig:pred_true_mag}. Again, an overall satisfactory agreement was obtained, with the majority of the predictions close to or symmetrically distributed around the ground truth value. Note that due to the distortion of the logarithmic function, the same deviation from ground truth at large and small magnitudes will look asymmetric and ``biased" towards smaller values. Therefore, two red dotted lines with constant deviation of $10^{-2}$ (outer) and $10^{-3}$ (inner) have been added in \Cref{fig:pred_true_mag}(a). Even at a large magnitude where the error of the ML model is also the largest, the difference in field vector magnitude is almost guaranteed to be smaller than $10^{-2}$. At a small magnitude, the difference in field vector magnitude is most likely to be smaller than $4 \times 10^{-3}$ and would typically be around $10^{-3}$. We did not notice any bias in our ML prediction results. The ML-predicted vectors are also very closely aligned with ground truth field vectors. As shown in  \Cref{fig:pred_true_mag}(b), most vectors would have an angle smaller than 10$^\circ$, and it is almost impossible to find a predicted vector with a more than 30 degrees angle from its ground truth counterpart.

\subsection{Dynamical Benchmark}

In addition to accurate predictions of spin torques, another important benchmark is whether the trained ML model can also faithfully capture the dynamical evolution of the itinerant spin model. To this end, we integrated the trained CNN model into the LLG dynamics simulations and compared the results with LLG simulations based on KPM~\cite{weisse06,wang18}. We consider simulations of a thermal quench process where an initially random magnet is suddenly quenched to nearly zero temperature at time $t = 0$. While our trained CNN model produces fairly accurate spin torques, small prediction errors still persist, as discussed in the previous section. Statistically, these prediction errors are similar to the stochastic noise $\bm\tau_i(t)$ in the LLG equation~(\ref{eq:LLG}). These site-dependent fluctuating random torques are similar to the thermal forces in Langevin dynamics. Both random forces are physically due to thermal fluctuations through coupling to a thermal bath at a fixed temperature. As a result, while the temperature of the ML-LLG simulations was set to exactly zero, a very low yet nonzero temperature $T = 0.001$ was introduced in the exact LLG dynamics to mimic the prediction error. 

%A dissipation constant $\alpha = 0.05$ was used for both simulations. Finally, we employ a second-order semi-implicit scheme~\cite{mentink10} to integrate the LLG equation. 

The model parameters of the s-d Hamiltonian are chosen to stabilize a spontaneous SkL ground state. Importantly, the emergence of skyrmion crystal not only breaks the spin-rotation symmetry but also breaks the lattice translational symmetry. The periods of this spatial modulation, \textit{i.e.}, the lattice constant of the skyrmion lattice, are determined by the underlying electron Fermi surface. Indeed, while an SkL state can be intuitively thought of as a periodic array of particle-like spin-textures, physically SkL phases often result from an instability caused by quasi-nesting of the electron Fermi surface that gives rise to a multiple-$Q$ magnetic order~\cite{ozawa16,hayami21,ozawa17}. 

In our case, the geometry of the Fermi surface at the chemical potential $\mu = -3.5$ allows significant segments to be connected by three wave vectors $\mathbf Q_1 = (\pi / 3a, 0)$ and $\mathbf Q_{2, 3} = \mathcal{R}_{\pm 2\pi/3} \cdot \mathbf Q_1$, related to each other by symmetry operations of the $\mathrm{D}_6$ group. Here $a$ is the lattice constant of the underlying triangular lattice. This means that maximum energy gain through electron-spin coupling is realized by spin helical orders with one of the above three wave vectors. Further analysis shows that the electron energy is further lowered by the simultaneous ordering of all three wave vectors, giving rise to an emergent triangular lattice of skyrmions. 

\begin{figure}[tb]
    \centering
    \includegraphics[width=\columnwidth]{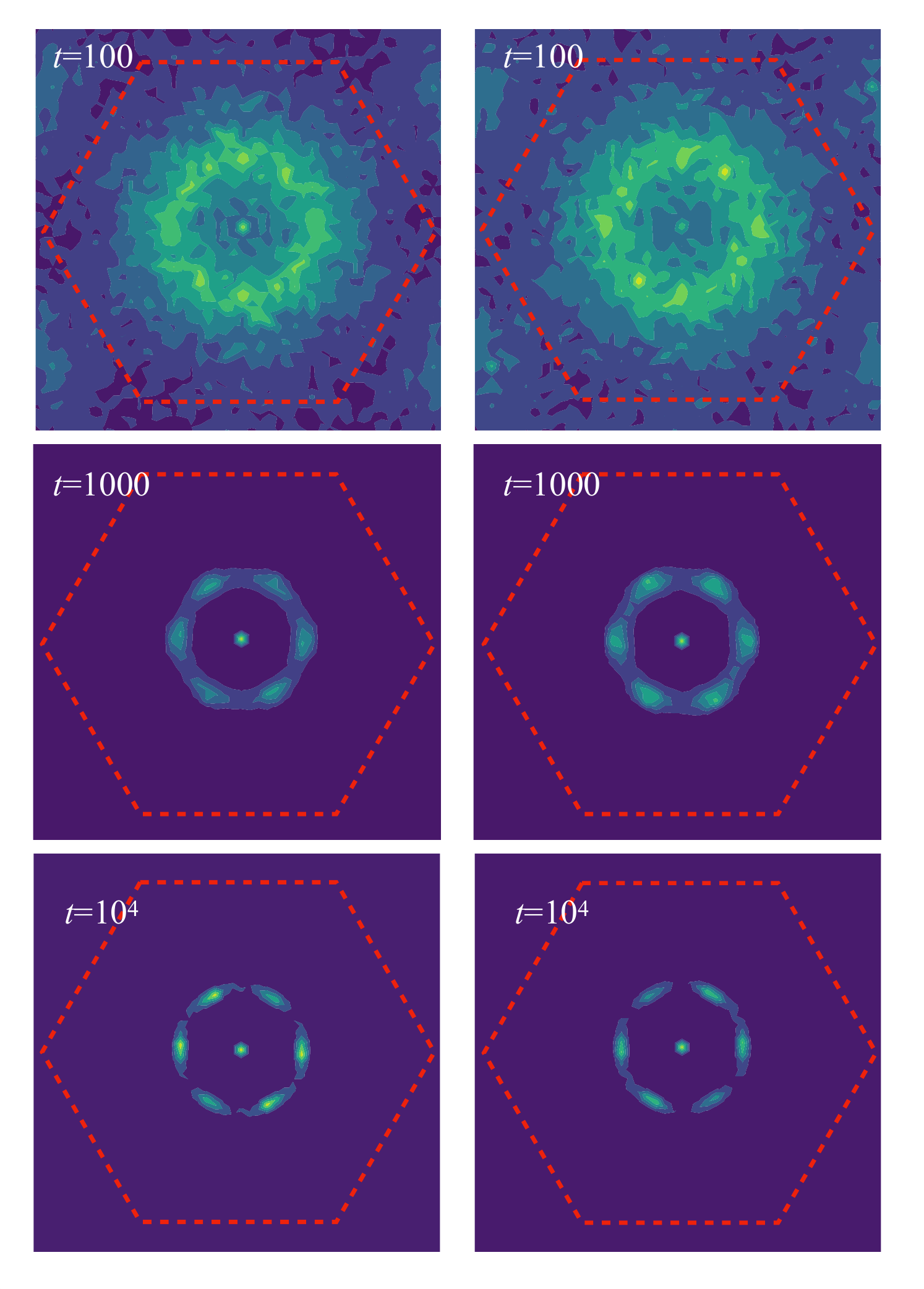}
    \caption{A comparison of spin structural factors obtained by averaging 30 independent LLG simulations based on KPM (left) and ML model (right). The same set of random initial conditions on a $48\times 48$ triangular lattice were used in both simulations. The red dashed lines indicate the first Brillouin zone of the momentum space. }
    \label{fig:ssf_comparison}
\end{figure}

The relaxation of the magnet after the thermal quench is dominated by the formation of the triangular SkL. A perfect SkL is distinguished by six Bragg peaks at $\mathbf q = \pm \mathbf Q_1$, $\pm \mathbf Q_2$, and $\pm \mathbf Q_3$ in momentum space. Yet, since the spin interactions are local in nature, the crystallization of skyrmions is inherently an incoherent process. Small crystallites of skyrmions are nucleated randomly separated by large domains of disjointed structures. To quantitatively characterize this crystallization process, we compute the time-dependent spin structure factor, which is defined as the square of the Fourier transform of the spin field
\begin{eqnarray}
	\mathcal{S}(\mathbf q, t) = \biggl\langle  \biggl| \frac{1}{N} \sum_{i=1}^N \mathbf S_i(t) \exp(i \mathbf q \cdot \mathbf r_i) \biggr|^2 \biggr\rangle,
\end{eqnarray}
where the bracket $\langle \cdots \rangle$ indicates averaging over thermal ensemble as well as initial conditions. The structure factor is itself the Fourier transform of the spin-spin correlation function in real space and can be directly measured in neutron scattering experiments. The spin structure factors at various times after the quench, obtained from LLG simulations based on both KPM and ML models, are shown in \Cref{fig:ssf_comparison}. Due to the stochastic nature of such simulations, the results are obtained by averaging 30 independent runs. Overall, the results from LLG simulations with the trained CNN model agree very well with those based on the numerically exact KPM. 

\begin{figure}[tb]
    \centering
    \includegraphics[width=\columnwidth]{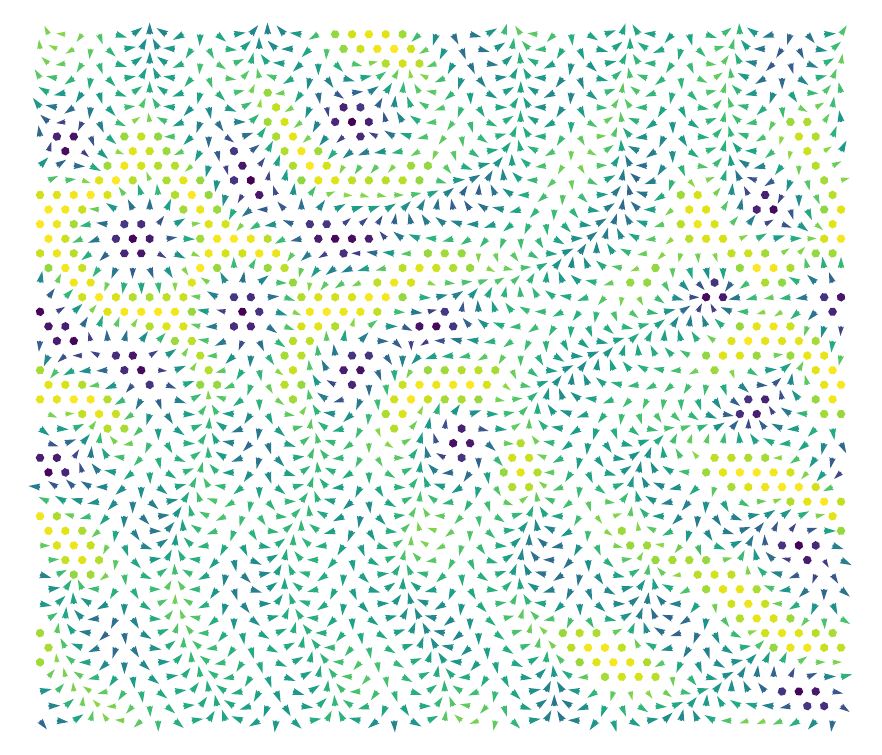}
    \caption{Snapshot of the spin configuration at the end of the LLG simulation with random initial conditions on a 48$\times$48 triangular lattice.}
    \label{fig:spin-config-quench}
\end{figure}

Both simulations show that a ring-like structure quickly emerges in the structure factor  after the quench. The radius of the ring is close to the length of the three nesting wave vectors~$\mathbf Q_{\eta}$, indicating the initial formation of skyrmions. As the system relaxed towards equilibrium, the ring-like structure becomes sharper. Moreover, the spectral weight starts to accumulate at the six spots corresponding to the $\pm \mathbf Q_\eta$ wave vectors. Physically, the emergence of the six broad segments corresponds to the growth of domains of the skyrmion lattice. The size of these intermediate skyrmion crystallites can be inferred from the width of the six spots. However, both simulations found that even at a late stage of the equilibration, the structure factor exhibits only six diffusive peaks at the nesting wave vectors, instead of sharp Bragg peaks as expected for a perfect SkL. The broad peaks at a late stage of the phase ordering thus indicate an arrested growth of SkL domains in real space. An example of the real-space spin configuration at $t = 10^4$ after the quench is shown in Fig.~\ref{fig:spin-config-quench}. The snapshot shows rather small triangular clusters of skyrmions coexist with stripe-like structures of different orientations, These stripes or helical spins corresponds to the single-$Q$ magnetic order which are meta-stable states of the s-d model.

This intriguing freezing phenomenon can be partly attributed to the frustrated electron-mediated spin interactions. Another important source is related to the degeneracy between skyrmions of opposite vorticity, or circulation of the in-plane spins. The two opposite circulations correspond to the topological winding number $w = \pm 1$ for the skyrmions. As discussed above, the spin-rotation symmetry is decoupled from the lattice in the s-d Hamiltonian~(\ref{eq:H_sd}), which provides a minimum model for centrosymmetric itinerant magnets without spin-orbit coupling. As a result, skyrmions with clockwise circulation is energetically degenerate with counter-clockwise ones.  This also means that SkL domains of the two opposite circulations are nucleated with roughly the same probability after the thermal quench. The subsequent annihilation of skyrmions with opposite vorticity thus prohibits the growth of a large coherent SkL.

%Since the final goal of developing our ML model is to enable large-scale simulation due to its superior speed, the predicted field vector must also be able to achieve good results in a dynamical benchmark. In this subsection, we present such a test case, where we replaced the exact field calculation with our machine learning model, ran an Atomistic Spin Dynamics (ASD) simulation with a temperature $T=0$ and a dampening factor $\alpha=0.05$ and compared the simulation with same initial spin configurations with exact field calculation. Both simulations start with the same random initial spin configuration and the spin structural factors after certain amount of steps are compared. We adopted the integration method of semi-implicit scheme, proposed in \citet{mentink2010}. The resulting structural factor is then averaged over 30 simulation runs with the same initial spin configuration, as there is inherent randomness in the simulation process. In order to maintain numerical stability, the time step of the simulations using the ML model is halved compared to that with the exact field calculation.

%As we can seen in \Cref{fig:ssf_comparison}, at $t=10000$, the same 6 segmented ring feature shows up in both simulations, indicating that our ML model is capable of achieving high enough accuracy for use in a simulation.

\begin{figure*}
    \centering
    \includegraphics[width=\textwidth]{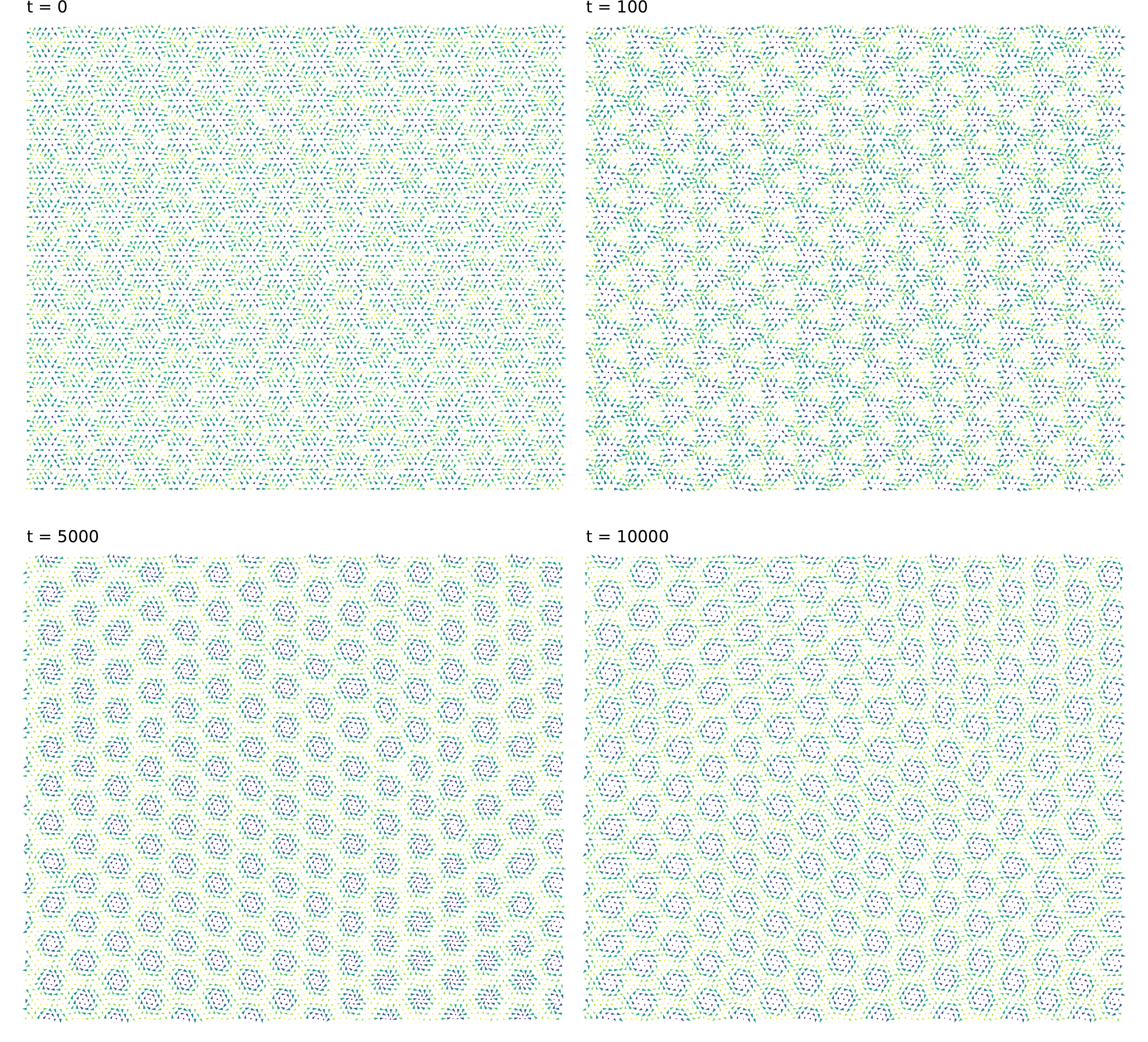}
    \caption{CNN-based LLG simulation on a $96\times 96$ lattice showing the restoration of a perturbed SkL. The CNN model was originally trained on a $48 \times 48$ lattice. The initial spin configuration is given by the SkL ansatz~(\ref{eq:skl}) with additional site-dependent random phases and amplitudes of $S_z$.}
    \label{fig:scalability}
\end{figure*}

\subsection{Scalability and Large-scale Simulation}

As discussed in Sec.~\ref{sec:architecture}, thanks to the locality property and the fixed-size kernels, the CNN model can be directly scaled to larger lattice systems without retraining, thus enabling large-scale dynamical simulations that are beyond conventional approaches. Here we demonstrate the scalability of the CNN spin-torque model by applying it to LLG simulations of large-scale SkL phases. Specifically, we perform LLG simulations of a perturbed SkL state on a $96\times 96$ lattice using a CNN model trained from simulations of a $48\times 48$ lattice.  As discussed above, the triangular skyrmion lattice, characterized by the three nesting wave vectors, can be viewed as a superposition of three helical spin orders. Explicitly, a perfect SkL can be approximated by the following ansatz~\cite{ozawa17,hayami21}
\begin{eqnarray}
	\label{eq:skl}
	& & \mathbf S_i \sim  \left( \cos\mathcal{Q}_{1i} - \frac{1}{2} \cos\mathcal{Q}_{2i} - \frac{1}{2} \mathcal{Q}_{3i} \right) \hat{\mathbf e}_1 \nonumber \\
	& & \qquad + \left( \frac{\sqrt{3}}{2} \cos\mathcal{Q}_{2i} - \frac{\sqrt{3}}{2} \cos\mathcal{Q}_{3i} \right) \hat{\mathbf e}_2  \\
	& & \qquad +  \left[ A \left(\sin\mathcal{Q}'_{1i} + \sin\mathcal{Q}'_{2i} + \sin\mathcal{Q}'_{3i} \right) + M \right] \hat{\mathbf e}_3, \nonumber
\end{eqnarray}
where $\hat{\mathbf e}_{1, 2, 3}$ are three orthogonal unit vectors, $\mathcal{Q}_{\eta i} = \mathbf Q_\eta \cdot \mathbf r_i$, and $\mathcal{Q}'_{\eta, i} = \mathcal{Q}_{\eta, i} + \phi$ are phase factors of the three helical orders, $\phi$, $A$, and $M$ are fitting parameters. To demonstrate that the ML model can indeed stabilize the SkL, which is the ground state of our chosen s-d Hamiltonian, we initialize the system with a perturbed array of skyrmions as shown in \Cref{fig:scalability}(a). The randomness in the initial state was introduced by allowing site-dependent parameters $\phi_i$, $A_i$ and $m_i$, which are randomly generated, in the above SkL ansatz~(\ref{eq:skl}). Contrary to completely random spins for the initial states in the previous dynamical benchmark, this initial state preserves a coherent structure of skyrmion winding numbers. As these topological numbers have to be conserved, the relaxation of the system is free of random annihilation of skyrmions. As shown in \Cref{fig:scalability}, our ML-based LLG simulations indeed find that a nearly perfect SkL is restored and stabilized over a long period of simulation time.

\begin{figure}
    \centering
    \includegraphics[width=\columnwidth]{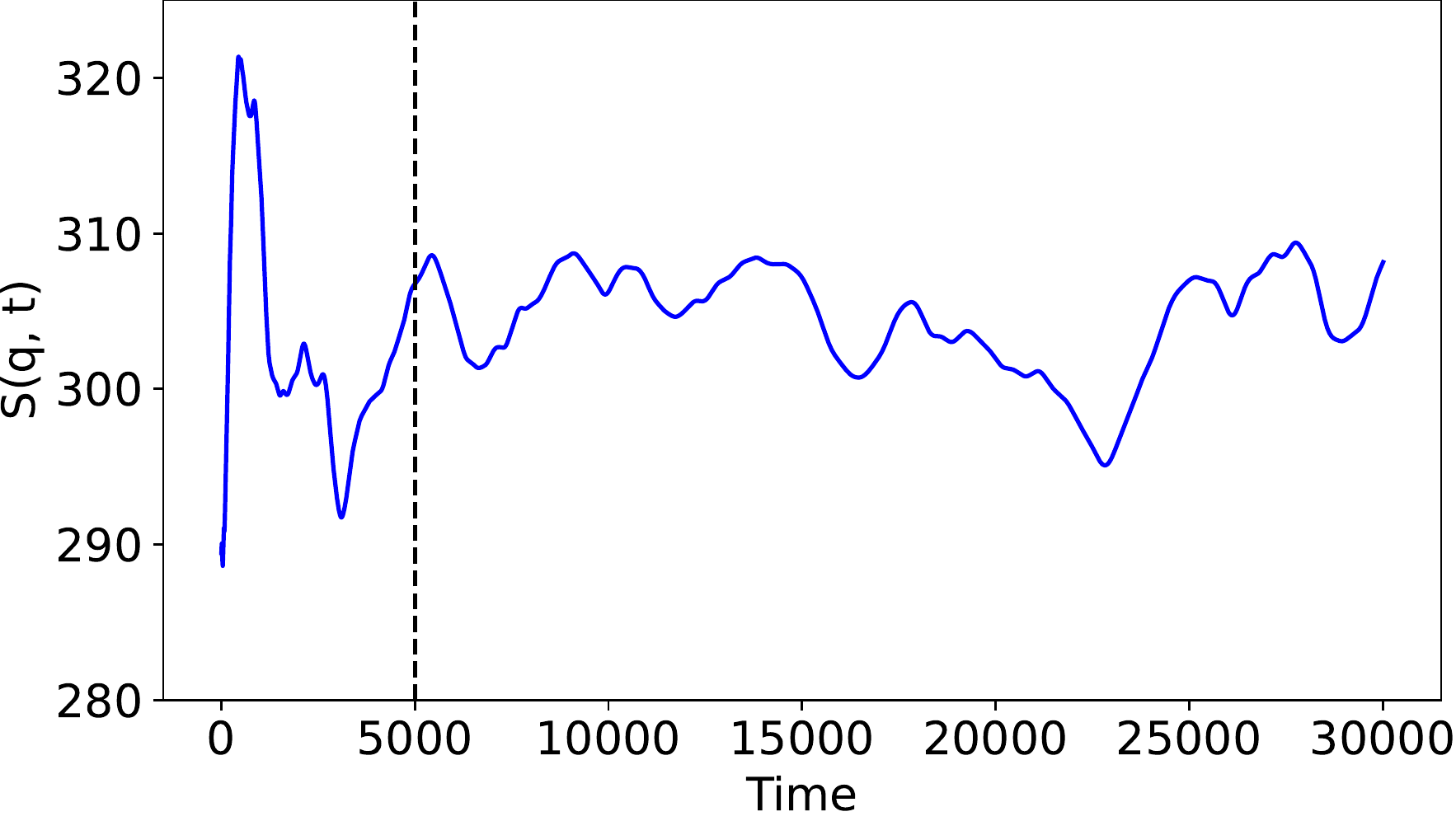}
    \caption{The evolution of structural factor of a much longer than training simulation with time, using perturbed Skyrmion initial condition. The dashed black line indicates the duration of the training simulation, and our ML-based LLG simulation is capable of keeping the structural factor constant for more than 5 times the duration of the training simulation.}
    \label{fig:ssf_time}
\end{figure}

We further investigate the scalability in the time domain by running our ML-based LLG simulation long past the duration of the training simulations. \Cref{fig:ssf_time} shows a roughly constant structural factor long beyond the duration of simulation snapshots used in training. While we noticed a huge decrease in structural factor between the time of 15,000 and 23,000, the structural factor quickly rebounds to its original stable value ($\mathcal{S}(\mathbf q, t)\approx$ 305). These temporal fluctuations can be ascribed to the prediction errors of the ML model. Yet, as also discussed above, such errors play a role similar to the stochastic noise in Langevin-type dynamics simulations. Our results thus demonstrate the robustness of SkL under small random perturbations. Importantly, this further benchmark highlight the scalability of our ML models not only in spatial domains (larger lattices), but also in temporal scales (much longer simulation times) as well.

%Lastly we demonstrate our ML model is scalable to enable simulations with larger lattice grids. In theory, a completely convolutional model, such as the one we constructed, is capable of scaling very well as long as the resolution of the input lattice does not change. Here, we showcase this point with a 96$\times$96 perturbed Skyrmion. Even though the ML model has never been trained on anything other than 48$\times$48 lattices, it is still capable of forming and maintaining stable Skyrmion vortex in a never-seen-before and larger lattice.

\subsection{Symmetry Requirements}
In order to incorporate the underlying symmetries of a physical system into an ML model, one needs to introduce appropriate biases (prior knowledge) through the statistical learning process. Two of the major approaches to this end are: i) \textit{data augmentation} based on the symmetry group of the system; ii-a) constructing \textit{symmetry-invariant} descriptors, or ii-b) constructing \textit{equivariant} neural network architectures \textit{w.r.t.} the symmetry group. These two types of approaches correspond to introducing the \emph{observational} and the \emph{inductive} biases, respectively, in the context of physics-informed machine learning literature (see, \textit{e.g.}, \cite{karniadakis2021physics,bronstein2017geometric,chen2020group,Nguyen2023parc,nguyen2023jcise,nguyen2023pep}). As discussed in Section~\ref{subsec:training}, the \emph{local} symmetries of our system, \textit{i.e.}, the spin-space and the real-space lattice symmetries \textit{a.k.a.} the internal (gauge) and the spacetime symmetries~\cite{laszlo2017unification}, consist of $G$-valued fields over the underlying lattice, where $G = \mathrm{SO} (3) \times \mathrm{D}_6 \,$. In the present work, we adopted the data augmentation approach as the mean to enforce the symmetry constraint for the reasons justified as follows.

First, we briefly summarize the theoretical justification of how data augmentation during our training phase is injecting the above-mentioned symmetries into the underlying supervised learning process (see~\cite{chen2020group,dao2019kernel,wang2022data} for details). To avoid cumbersome notation, we denote a pair of spin configurations and its corresponding torque field $(\mathbf{S}, \mathbf{T}) = (\{\mathbf{S}_i\} \, , \, \{\mathbf{T}_i\})$ by $\mathbf{F}$. Our training data $\mathbf{F}_1 , \cdots , \mathbf{F}_n$ consist of independent identically-distributed (i.i.d.) samples from a probability distribution $\mathbb{P}$ over the space of all spin-torque fields. It is of fundamental importance that the probability distribution $\mathbb{P}$ remains \emph{invariant} under the \emph{action} of each local symmetry $g \in \mathcal{G} \,$, where $\mathcal{G}$ denotes the space of all local symmetries of the system\footnote{The action of a local symmetry $g \in \mathcal{G}$ over a spin $\mathbf{S}$ and a torque $\mathbf{T}$ field is the induced transformation by $g\,$. We denote it by $g \cdot \mathbf{F} := (g \cdot \mathbf{S} \, , \, g \cdot \mathbf{T})$.}. Hence, the data augmentation process can be considered as enriching our set of samples from the probability distribution $\mathbb{P}$, where our goal is to learn it, by adding transformed spin-torque fields $g \cdot \mathbf{F} , \, g \in \mathcal{G}$.

During the training procedure, at each step $t$, a minibatch $B_t$ of spin-torque
$(\mathbf{S}, \mathbf{T})$ samples of size $|B_t|$ is chosen, and a random local symmetry $g_{t,b} \in \mathcal{G}$ is applied to each spin $\mathbf{S}_b$ and torque $\mathbf{T}_b$ field, $b \in B_t$. Then, according to the stochastic gradient descent (SGD) algorithm, the parameters $\theta$ of the CNN model $f_\theta$ get updated as
\begin{equation}
    \theta_{t+1} = \theta_t - \frac{\eta}{|B_t|} \sum_{b \in B_t} \nabla_\theta L \left( {f_\theta (g_{t,b} \cdot \mathbf{S}_b) \, , \, (g_{t,b} \cdot \hat{\mathbf{T}}_b}) \right) \, ,
\end{equation}
where $L$ denotes the loss function given by Eq.~(\ref{eq:loss}), and $\eta$ is the learning rate. In other words, the augmented SGD can be considered as minimization of the \emph{empirical risk} associated with the following \emph{augmented loss function}
\begin{equation}
    \int_{\mathcal{G}} L \left( {f_\theta (g \cdot \mathbf{S}) \, , \, (g \cdot \hat{\mathbf{T}}}) \right) \, d \mathbb{Q}(g) \, ,
\end{equation}
in which one takes an average along the whole orbit of the group action w.r.t. a probability distribution $\mathbb{Q}$ over $\mathcal{G}$. It can be proved that data augmentation based on the underlying symmetry group reduces the variance of general estimators and improves their generalizability~\cite{chen2020group}. 

\begin{figure}
    \centering
    \includegraphics[width=\columnwidth]{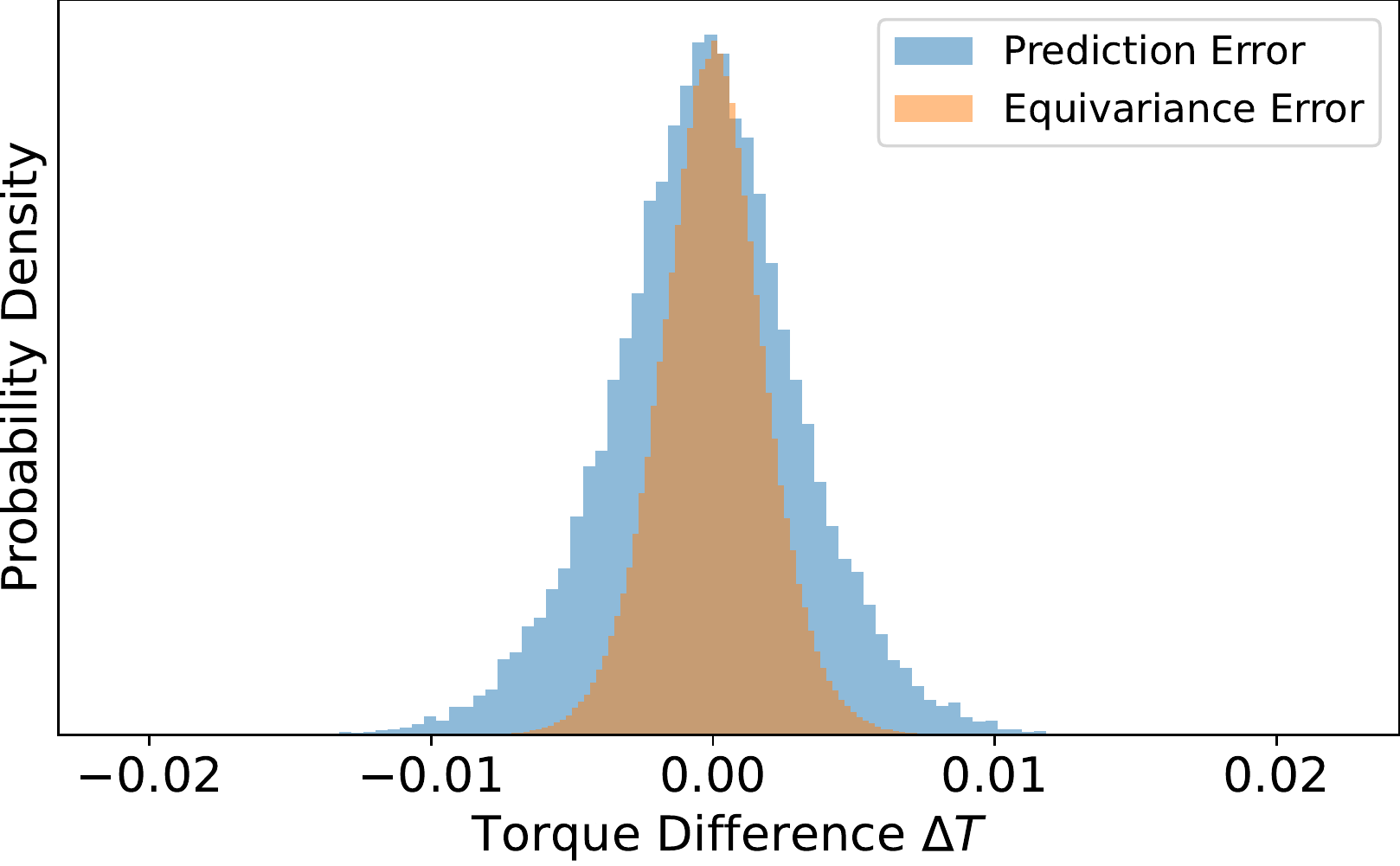}
    \caption{Distribution of the equivariance error $err_{eq} := f_\text{CNN}(\{R\mathbf{S}_i\}) - Rf_\text{CNN}(\{\mathbf{S}_i\})$ where $R$ is an arbitrary rotation. The overall prediction error (`predicted torque' minus `ground truth torque') on the test dataset is superimposed for comparison. 
    The equivariance error is sufficiently smaller than the overall prediction error, implying that the model can preserve the underlying symmetry of the physics system reasonably well.}
    \label{fig:equivariance_error}
\end{figure}

The above theoretical justification can further be validated through empirical data. \Cref{fig:equivariance_error} shows the typical prediction error (blue), the difference between predicted and ground truth torque, and the equivariance error (orange), defined as $f_\theta (g \cdot \mathbf{S}) - g \cdot f_\theta (\mathbf{S})$. As can be seen in the figure, the equivariance error is smaller than the typical prediction error, indicating that in practice data augmentation employed is capable of preserving the underlying symmetry of the physics system to a satisfactory degree.
%({\color{red}We could compare data augmentation versus equivariant CNN. Xinlun may need to do some experiments.})
%{\color{red}
%\begin{enumerate}
%    \item Accuracy plot showing data augmentation actually gives more accurate results.
%    \item Predicted torque under different rotations. Similar to the animation (the 'input - feature map - stabilized view' figure) in \url{https://github.com/QUVA-Lab/e2cnn}
%\end{enumerate}
%}

\subsection{Locality Principle and Receptive Fields}

To attest to the locality principle, we analyze the receptive field of our CNN model in this section. As discussed in Section~\ref{sec:convnet}, the receptive field of a convolution layer $f_m$ is defined to be the support $\supp (h_m)$ of the corresponding convolution kernel $h_m$, \textit{i.e.}, the region where the function values of $h_m$ are nonzero tensors. The receptive field of the entire CNN model is computed as the Minkowski sum of the receptive fields of individual convolution layers, or $\text{RF} = \text{supp}(h_1) \oplus \cdots \oplus \text{supp}(h_L)$. For our model, in which there are 10 layers in depth with each layer comprised of $5 \times 5$ convolution kernels of stride 1, the size of the receptive field is calculated to be 41. This implies that, in principle, the spin directions of the 41-neighborhood can influence the prediction of the torque at the lattice position $i$.

\begin{figure}
    \centering
    \includegraphics[width=\columnwidth]{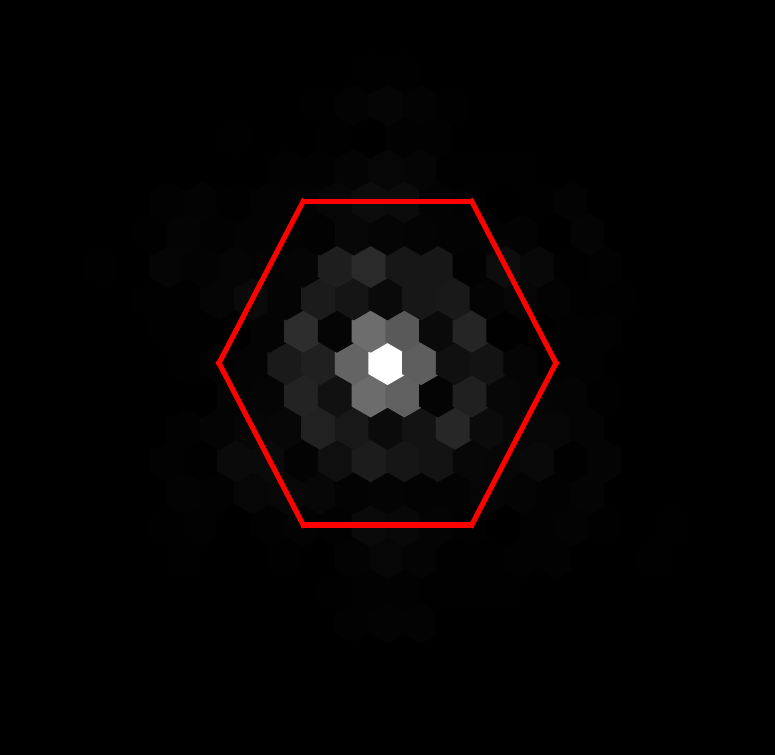}
    \caption{The effective receptive fields (ERFs) of the ML model. Since for each lattice site both the input and output tensors have 3 components, there are in total 9 ERFs corresponding to the 9 partial derivatives of the 3 output torque components with respect to the 3 input spin components. The sum of these absolute values of these derivatives are then presented in this figure, with blacker pixels indicating smaller derivative value. The red line roughly traces the non-zero value regions of the ERF. }
    \label{fig:locality}
\end{figure}

However, the na\"ive computation of the receptive field size may be misleading because the kernel size of a convolution layer simply just indicates the theoretical maximum of the receptive field. On the other hand, the actual region of nonzero values can be much smaller than the theoretical receptive field size. To this end, we used the approach of \citet{Luo2017} to compute the \textit{effective} receptive field size, in which function values are practically nonzero. Figure~\ref{fig:locality} shows the result of such a calculation performed on the trained CNN model. The red hexagonal line delineates the region inside the hexagon where function values are practically nonzero and the region outside the hexagon where function values are practically zero. The grayscale values inside the hexagon indicate different levels of influence of neighboring spins in computing the torque vector. As can be seen from the figure, at the lattice location $i$, which is the center of the red hexagon, the weighting factor is the largest, implying that $\mathbf{T}_i$ is predominantly determined by $\mathbf{S}_i$. The 1-neighborhood $\mathcal{N}_1(i)$, \textit{i.e.}, the immediate neighbors to the lattice location $i$, also has bright intensity values, implying that the relative configuration of the spin direction $\mathbf{S}_i$ to its neighboring spins $\mathbf{S}_j$ at $j\in \mathcal{N}_1(i)$ also have a significant influence to the output torque $\mathbf{T}_i$. Similarly, it appears that the spin directions of the 3-neighborhood have strong influences on torque prediction, while small influence can be detected all the way to 6-neighborhood.

This result is consistent with previous ML spin-torque models based on symmetry-invariant descriptors~\cite{zhang20,zhang21}, which shows that the spin dynamics of similar s-d models can be nicely captured by BP-type models based on fully connected NN with input from a neighborhood up to $r_c \sim 5$ lattice constants. Physically, as discussed above, the finite sizes of effective receptive field are due to the locality nature of the spin torques. However, the range of locality can only be indirectly determined from exact calculations. In practice, the cutoff radius is treated as an ad hoc parameter in BP-type ML models, or is determined through trial and error. It is thus worth noting that the CNN model offers a systematic and rigorous method to determine this important physical attribute of electronic models.

\section{\label{sec:conclusion}Conclusion and outlook}

In this paper, we presented a CNN model to predict spin torques directly from input spin configuration for large-scale LLG dynamics simulations of itinerant magnets. Our CNN model is purely convolutional without any fully connected (dense) layers, and thus presents a distinct advantage of built-in locality. Central to each CNN layer is the convolution with a kernel or filter, which can be viewed as a Green's function representing finite responses to a local source. As each kernel is characterized by a finite set of trainable parameters, the CNN model can be used for dynamical simulations on larger system sizes without rebuilding or retraining the neural network. We demonstrated our ML approach on a triangular-lattice s-d model which exhibits a skyrmion crystal in its ground state. Using the ML-predicted torques in the LLG dynamics simulations, we showed that the trained CNN model can successfully reproduce the relaxation of the skyrmion phase of the itinerant spin models. We further demonstrated the scalability and transferability of our approach by showing that large-scale LLG simulations based on our CNN model are able to stabilize a perturbed skyrmion lattice and maintain it for a long period of time.

Contrary to the ML force-field models based on the Behler-Parrinello scheme, our CNN model directly predicts torques, which are the spin analogue of atomic forces. In BP-type approaches, ML models, either Gaussian process regression or fully connected neural nets, are built to predict local energy, which cannot be directly compared with exact calculations. The forces are obtained from derivatives of the total energy, which is the sum of all local energies. The introduction of local energy takes advantage of the locality property and also facilitates the incorporation of symmetry into the ML models. Yet, the fact that forces are computed indirectly from derivatives of energy also restricts BP-type models to the representation of conservative forces and quasi-equilibrium electron systems. On the other hand, our CNN approach can be used to describe both conservative as well as non-conservative spin torques. This capability is particularly important for ML modeling of out-of-equilibrium driven systems where the electron-mediated torques are non-conservative. A representative example is the spin transfer torque which plays an important role in spintronics applications.

For future work, we are currently looking into ways to enforce constraints due to either symmetry or conservation laws more strictly and rigorously. To this end, previous computer vision literature on equivariant CNNs (see \textit{e.g.}, \citet{Geiger2022}) may shed light on how to constrain CNN layers to preserve $\mathrm{SO} (3)$ and $\mathrm{D}_6$ symmetries. Moreover, the present work is limited to approximating torques using spin directions at each time step and does not provide a direct solution to the LLG equation in Eq.~\ref{eq:LLG}. In a recent body of literature, however, there have been attempts to solve governing partial differential equations (PDE) of physics directly using so-called physics-aware deep neural networks (see \textit{e.g.}, Nguyen \textit{et al.}~\cite{Nguyen2023parc}). Using these physics-aware CNN methods, we expect to attain faster and more accurate approximations of the spin dynamics, which is going to be another meaningful direction of research.

%We constructed a convolutional neural network design to predict local exchange field directly from input spin configuration, leveraging the superior speed, built-in locality and built-in scalability of a CNN model. The ML model achieves great accuracy in all three field components, field magnitude and directions. The ML model is also capable of replacing the traditional field calculation method in a numerical simulation and reproducing faithful relaxation of the input system. We also show that the scalability and transferrablitiy of the ML model by stabilizing a Skyrmion lattice that is larger than the ones in the training set.

\begin{acknowledgments}
SZ and GWC acknowledge the support from the US Department of Energy Basic Energy Sciences under Contract No.~DE-SC0020330. 
SSB and PCHN were supported by the National Science Foundation under Grant No.~DMREF-2203580. XC acknowledge the support from the Jefferson Scholars Foundation and the benefactor of the Edward P. Owens Jefferson Fellow.
\end{acknowledgments}

%\appendix

%\section{Appendixes}

% The \nocite command causes all entries in a bibliography to be printed out
% whether or not they are actually referenced in the text. This is appropriate
% for the sample file to show the different styles of references, but authors
% most likely will not want to use it.
\nocite{*}

\bibliography{apssamp}

\end{document}